\documentclass[preprint,showpacs,preprintnumbers,amsmath,amssymb]{revtex4} 
 
\usepackage{graphicx}
\usepackage{dcolumn}
\usepackage{bm}

\begin{document} 
 
\bibliographystyle{apsrev}

\preprint{Atomic charge distribution ... }   
   
\title{Atomic-charge distribution in glasses by terahertz spectroscopy}

\author{S.N. Taraskin}   
 \email{snt1000@cam.ac.uk}   
\affiliation{St. Catharine's College and Department of Chemistry, University of Cambridge,   
             Cambridge, UK}   
 
\author{S.I. Simdyankin}   
\email{sis24@cam.ac.uk}   
\affiliation{Department of Chemistry, University of Cambridge,   
             Cambridge, UK}   
 
\author{S.R. Elliott}   
\email{sre1@cam.ac.uk}   
\affiliation{Department of Chemistry, University of Cambridge,   
             Cambridge, UK}

\date{\today}
   
\begin{abstract}  
It is demonstrated that the width of the uncorrelated atomic-charge
distribution in glasses can be extracted 
from the frequency dependence of the coupling coefficient 
for the far-infrared absorption measured
experimentally by the time-domain terahertz spectroscopy technique. 
This value for As$_2$S$_3$ glass is
found to be $0.12$ (e). 
A density functional theory-based tight-binding molecular dynamics model
of As$_2$S$_3$  glass qualitatively supports these findings.
\end{abstract}   
   

   
\maketitle   
   
  
\section{Introduction}  
\label{s0} 
 
Disordered solids, such as glasses, are characterized by a lack of structural 
order, which leads to disorder in many physical quantities.  
For example, atomic charges in glasses fluctuate in space.  
The origin and properties of such fluctuations are of considerable interest in 
the field of disordered systems.  
One such intriguing question concerns the scale of charge fluctuations  
and possible charge ordering in glasses 
\cite{Pasquarello_97,Massobrio_04,Blaineau_04,Giacomazzi_06,Giacomazzi_07}.  
Another intriguing question is related to the possibility of obtaining 
information about the charge distribution from experiment.  
It has been recently discussed how  terahertz absorption spectroscopy can 
be used for this purpose \cite{Taraskin_06:PRL}.   
   
In this paper, we study the atomic charge distributions in glassy arsenic 
sulfide, As$_2$S$_3$,  by means of first principle tight-binding molecular 
dynamics simulations and compare the characteristics of these distributions   
with those obtained from experimentally available far-infrared (FIR)
absorption  coefficient measurements.   
 
\section{Theoretical background}  
\label{s1} 
   
First, we recall how the absorption coefficient in the FIR region is related to the atomic
charge distribution  \cite{Taraskin_06:PRL}.     
The expression for the linear absorption coefficient of photons,
$\alpha(\omega)$, caused  
by the interaction with harmonic atomic vibrations in solids is given
by the following 
expression (see e.g.  \cite{Pasquarello_97}),  
\begin{equation} 
\alpha(\omega) = 
 \frac{2\pi^2n}{c\sqrt{\varepsilon_{\infty}}} 
\left\langle \sum_\alpha \left|  
\sum_{i\beta} \frac{Z_{i,\alpha\beta}}{\sqrt{m_i}}{e}_{i\beta}(\omega) 
\right|^2  
g(\omega) \right\rangle  
~,  
\label{e1}  
\end{equation} 
where $m_i$ and $Z_{i,\alpha\beta}$ are the mass and the dynamical charge  
tensor for atom $i$ ($i=1,\ldots,N$, with $N$ being the number of atoms in 
a solid of volume $V$; $\alpha$ and $\beta$ run over Cartesian
coordinates), 
 ${e}_{i\beta}(\omega)$ is the $\beta$-component of the  
vibrational eigenvector of frequency $\omega$ corresponding to atom $i$,  
$\varepsilon_{\infty}$ stands for the high-frequency dielectric constant,   
$n=N/V$ is the atomic concentration and angular brackets denote 
configurational averaging.  
As follows from Eq.~(\ref{e1}), the IR absorption coefficient is proportional to the self-averaging value of the
vibrational density of states (VDOS), 
\begin{equation} 
g(\omega)=(3N)^{-1}\sum_j \delta(\omega-\omega_j) 
~,  
\label{eq:VDOS}  
\end{equation} 
where $j$ runs over all the 
eigenfrequencies, i.e.  
\begin{equation} 
\alpha(\omega)= \left\langle C(\omega)\right\rangle  g(\omega) 
 ~.  
\label{eq:C_def}  
\end{equation} 
The coefficient of proportionality, 
$\left\langle  C(\omega)\right\rangle$, 
is called   
the coupling coefficient between IR photons and atomic vibrations for linear 
light absorption.  
The VDOS behaves universally ($g(\omega) \propto \omega^2$  
according to the Debye law) in the FIR regime and thus all interesting and 
possibly universal features can be attributed to the frequency dependence of 
the coupling coefficient.     
 
The absorption coefficient given by Eq.~(\ref{e1})  
is temperature independent and some variations of $\alpha(\omega)$ with  
temperature  found  
experimentally for microwave frequencies \cite{Strom_77},  
$\omega/2\pi c \alt 1~\text{cm}^{-1}$,  
which may  possibly be attributed to excitations of two-level systems and/or  
to highly anharmonic atomic modes, are not considered here.  
The dynamical charge tensors
are crucial for a correct description of the peak positions and their relative intensities in the
bulk of the vibrational band (above the FIR region)
 \cite{Wilson_96,Pasquarello_97} but  
are not so significant in the FIR regime  and thus a simpler  
(rigid-ion) model \cite{Maradudin_61} can be used, so that   
\begin{equation} 
C(\omega) = 
C_0  
\left|  
\sum_i \frac{q_i}{\sqrt{m_i}}{\bf e}_{i}(\omega) 
\right|^2  
~,     
\label{e2}  
\end{equation} 
with $q_i$ being the temporally fixed but spatially fluctuating atomic charges 
and   
\begin{equation}  
C_0=2\pi^2n/c\sqrt{\varepsilon_{\infty}}  
~.     
\label{eq:C0}  
\end{equation} 
In the well-studied case of ordered systems, where the charges do not
fluctuate, and the 
eigenmodes are phonons, the coupling coefficient is non-zero only for
optic modes at the 
centre of the Brillouin zone. In disordered systems, structural
disorder leads to charge 
transfer between atoms, i.e. to disorder in atomic charges $q_i$,
 and to intrinsic disorder in the
components of the eigenvectors which lose their translational
invariance. These two related 
sources of disorder, encoded in Eq.~(\ref{e2}),   
are responsible for the peculiar behaviour of $C(\omega)$ in amorphous systems,  
which actually has a universal functional form in the FIR regime  
\cite{Taraskin_06:PRL},  
\begin{equation} 
\left\langle C(\omega) \right\rangle \simeq A +B \omega^2 
~,  
\label{e0}  
\end{equation} 
where $A$ and $B$ are material-specific constants.  
 
In order to see this, we use two facts known about the structure  
of the eigenmodes in the FIR regime and about the  
distribution of atomic charges.  
First, the disordered eigenmodes in the FIR regime resemble plane waves  
\cite{Taraskin_00:PRB_IR1,Taraskin_98:PhilMag} characterized by 
pseudo-wavevectors $k$ and  exhibiting pseudo-dispersion,  
$\omega_j(k) = c_j k $ (with $c_j$ being the sound velocity for branch $j$)  
 and can be well approximated \cite{Taraskin_06:PRL} by a plane wave   
characterized by wavevector ${\bf k}$ and unit  
polarization vector ${\bf {\hat p}}_{\bf k}$,  
\begin{equation} 
{\bf e}_{i}(\omega) \simeq  
\sqrt{\frac{m_i}{N \overline{m}}} {\bf {\hat p}}_{\bf k}  
e^{\text{i}{\bf k}(\omega)\cdot{\bf r}_i} 
~,  
\label{eq:plane_wave}  
\end{equation} 
with  
$\overline{m}=N^{-1}\sum_i m_i$ and ${\bf r}_i$ being the position vector of 
atom $i$.   
 
The second useful fact concerns the distribution of charges in disordered 
systems.  
It has been found in simulations 
\cite{Pasquarello_97,Blaineau_04,Giacomazzi_06,Giacomazzi_07} 
that the charges  
in the models of some amorphous materials preserve approximately charge 
neutrality within certain structural units.  
For example, the SiO$_4$ structural units in vitreous silica are approximately 
electro-neutral \cite{Pasquarello_97} meaning that the positive charge
on an Si atom is approximately equal in magnitude to half 
of the sum of the charges on the four nearest oxygen atoms. The values
of silicon and oxygen 
charges vary strongly between structural units, depending on local
structural characteristics 
such as the  Si-O-Si bond angle. Moreover, the electro-neutrality
within the structural units 
is maintained only approximately (see below) and there is always a
stochastic component 
in the charge distribution due to intermediate and long-range fluctuations in the structure.
These observations allow the values of atomic charges to be split into
two components, $q_i=q_{1i} +q_{2i}$,  
with $q_{1i}(\{ {\bf r}_i\})$  representing uncorrelated charge components, and  
the random charges $q_{2i}$ satisfying local charge neutrality.  
 
The values of $q_{1i}(\{ {\bf r}_i\})$  
depend on the  atomic coordinates $\{ {\bf r}_i\}$ in  
a complicated fashion so that we can approximately assume the absence of  
correlations between $q_{1i}$ on different atoms, i.e.  
\begin{equation} 
\langle  q_{1i} q_{1j} \rangle \simeq  
\langle q_{1i}\rangle\langle q_{1j} \rangle \simeq  
\sigma_{1i}^2 \delta_{ij} 
~,  
\label{eq:q1_ij}  
\end{equation} 
where  
the variance $\sigma_{1i}^2$ can vary for different type of 
atoms, or for the same atoms but e.g. that are  abnormally
coordinated.
Similarly, we assume no correlations between $q_{1i}$ and atomic 
position vectors, so that 
\begin{equation} 
\langle q_{1i} e^{\text{i}{\bf k}\cdot{\bf r}_j} \rangle  
\simeq  
\langle q_{1i}\rangle  
\langle e^{\text{i}{\bf k}\cdot{\bf r}_j} \rangle \simeq 0 
~.   
\label{eq:q1_R}  
\end{equation} 

The random charges $q_{2i}$ obeying local charge neutrality  
can be imagined as resulting from charge transfers between
nearest-neighbour atoms, i.e.  
$q_{2i} = \sum_{j\ne i}\Delta q_{ji} $, where $j$ runs through all  
the nearest-neighbours of atom $i$ and $\Delta q_{ji}$  
($=-\Delta q_{ji}$) is the  
charge transfer from the originally neutral atom $j$ to the originally  
neutral atom $i$.  
In heteropolar crystals, the values of $\Delta q_{ji}$  
are not random and finite. 
In disordered systems, the values of $\Delta q_{ji}$ are distributed  
around mean value(s) which do not necessary coincide with those   
for crystalline counterparts (see e.g. \cite{Pasquarello_97}).  
Such fluctuations and deviations of means in $\Delta q_{ji}$ are due to   
distortions in local structural units, e.g. in bond angles and bond 
lengths.  
We also assume that there are no correlations between randomly 
fluctuating charges $q_{1i}$ and local charge transfers,  
\begin{equation} 
\langle  q_{1i} q_{2j} \rangle \simeq   
\langle q_{1i}\rangle\langle q_{2j} \rangle \simeq 0 
~.  
\label{eq:q1q2}  
\end{equation} 

The configurationally averaged coupling coefficient can be recast in terms 
of correlated and uncorrelated charges in the following manner,  
\begin{equation} 
 \tilde{C}(\omega) =\frac{ \langle C(\omega)\rangle \sqrt{\overline{m}}}{C_0} = 
\left\langle  
N^{-1}   
\left|  
\sum_i  q_{1i} e^{\text{i}{\bf k}\cdot{\bf R}_i} +  
\sum_i q_{2i} e^{\text{i}{\bf k}\cdot{\bf R}_i} 
\right|^2  
\right\rangle  
 \equiv  
\left\langle N^{-1}   
\left|  
S_1 + S_2 
\right|^2 
\right\rangle  
~,      
\label{e3}  
\end{equation} 
where $S_n=\sum_i  q_{ni} e^{\text{i}{\bf k}\cdot{\bf R}_i}$ 
($n=1,2$).  
In the absence of correlations between  $q_{1i}$ and $q_{2j}$  
(see Eq.~(\ref{eq:q1q2})), the above formula for $\tilde{C}(\omega)$ 
reduces to  
\begin{equation} 
 \tilde{C}(\omega) = N^{-1}   
\left( 
      \left\langle\left| S_1 \right|^2\right\rangle + 
      \left\langle\left| S_2 \right|^2\right\rangle  
\right) 
~.       
\label{eq:C_tilde}  
\end{equation} 
The first component in Eq.~(\ref{eq:C_tilde}) can be further 
simplified as,  
\begin{equation} 
 N^{-1}   
      \left\langle\left| S_1 \right|^2\right\rangle 
\simeq N^{-1} \sum_{ij} \langle q_{1i} q_{1j} \rangle  
\left \langle e^{\text{i}{\bf k}\cdot({\bf r}_{j}-{\bf r}_{i}) } 
\right\rangle \simeq   
N^{-1}\sum_i \sigma^2_{1i} = {\overline\sigma}^2_1 
~,        
\label{eq:S1}  
\end{equation} 
where we have used Eqs.~(\ref{eq:q1_ij})-(\ref{eq:q1_R}).  
In the case of a two-component system containing $N_1$ and $N_2$ atoms 
of different types,  
${\overline\sigma}^2_1 
= 
(N_1/N)\sigma^2_{11} +(N_1/N)\sigma^2_{12} $.  
Therefore, the first contribution in the coupling coefficient is 
frequency independent and depends only on the variance of uncorrelated 
charge distributions.  
 
The second component in the coupling coefficient,  
$N^{-1} \left\langle\left| S_2 \right|^2\right\rangle$,  
which is due to random and locally correlated  
charge fluctuations, does not contain the frequency-independent  
part and, in fact, is proportional to $\omega^2$.  
This can be demonstrated using the bond representation for $S_2$,  
\begin{equation} 
S_2 = \left\langle \sum_{(ij)}\Delta q_{ij}  
\left(  e^{\text{i}{\bf k}\cdot{\bf r}_j}-  
e^{\text{i}{\bf k}\cdot{\bf r}_i} 
\right)\right\rangle =  
 \left\langle 
\sum_{(ij)}\Delta q_{ij}  e^{\text{i}{\bf k}\cdot{\bf r}_{(ij)}}  
\left(  e^{\text{i}{\bf k}\cdot{\bf r}_{ij}/2}-  
e^{\text{i}{\bf k}\cdot{\bf r}_{ij}/2} 
\right)\right\rangle   
~,        
\label{eq:S2}  
\end{equation} 
with ${\bf r}_{ij}={\bf r}_{j}-{\bf r}_{i}$  
and  ${\bf r}_{(ij)}=({\bf r}_{j}+{\bf r}_{i})/2$, where the sum is 
taken over all the bonds $(ij)$ in the system.  
In the absence of the plane wave ({\bf k}=0), this sum  
equals zero, thus reflecting global charge neutrality  
of locally neutral units (the contribution from each bond  
is exactly zero due to the local charge neutrality).  
In the FIR regime, ${\bf k}\cdot{\bf r}_{ij} \ll 1$ and  
Eq.~(\ref{eq:S2}) can be recast as  
\begin{equation} 
S_2 \simeq k \sum_{(ij)}\Delta q_{ij}  e^{\text{i}{\bf k}\cdot{\bf r}_{(ij)}}  
\left(\text{i}{\hat{\bf n}}\cdot{\bf r}_{ij}\right) 
~,        
\label{eq:S2a}  
\end{equation} 
with ${\bf k} = k{\hat{\bf n}}$.  
Consequently, the contribution from the correlated charges  
to the coupling coefficient is  
\begin{equation} 
\langle\left| S_2 \right|^2\rangle  
\simeq k^2  
\left\langle  
\sum_{(ij)(i'j')} 
\Delta q_{ij}\Delta q_{i'j'}  e^{\text{i}{\bf k}\cdot({\bf r}_{(ij)}-{\bf r}_{(i'j')})}  
\left({\bf n}\cdot{\bf r}_{ij}\right) 
\left({\bf n}\cdot{\bf r}_{i'j'}\right) 
\right\rangle  
~,        
\label{eq:S2b}  
\end{equation} 
so that $\langle\left| S_2 \right|^2\rangle \propto k^2 \propto \omega^2$  
(implying linear dispersion in the FIR regime).  
The double sum in Eq.~(\ref{eq:S2b}) depends on 
precise structural details of the material but does not 
depend on $k$ in the FIR range ($\propto \text{Const} + 
\text{O}(k^2)$)  
and thus the $\omega^2$-dependence of 
the second contribution in the coupling coefficient is a general 
feature of the FIR absorption in disordered solids.  
 
Therefore, we have demonstrated that uncorrelated atomic charges
result in the frequency-independent part of the coupling coefficient,
while locally correlated charges, maintaining the 
charge neutrality within local structural units, give rise to the
quadratic frequency dependence of the coupling coefficient.

\section{Results}  
\label{s2} 
 
\begin{figure}[ht] %
\vskip2truecm  
\centerline{\includegraphics[width=13cm]{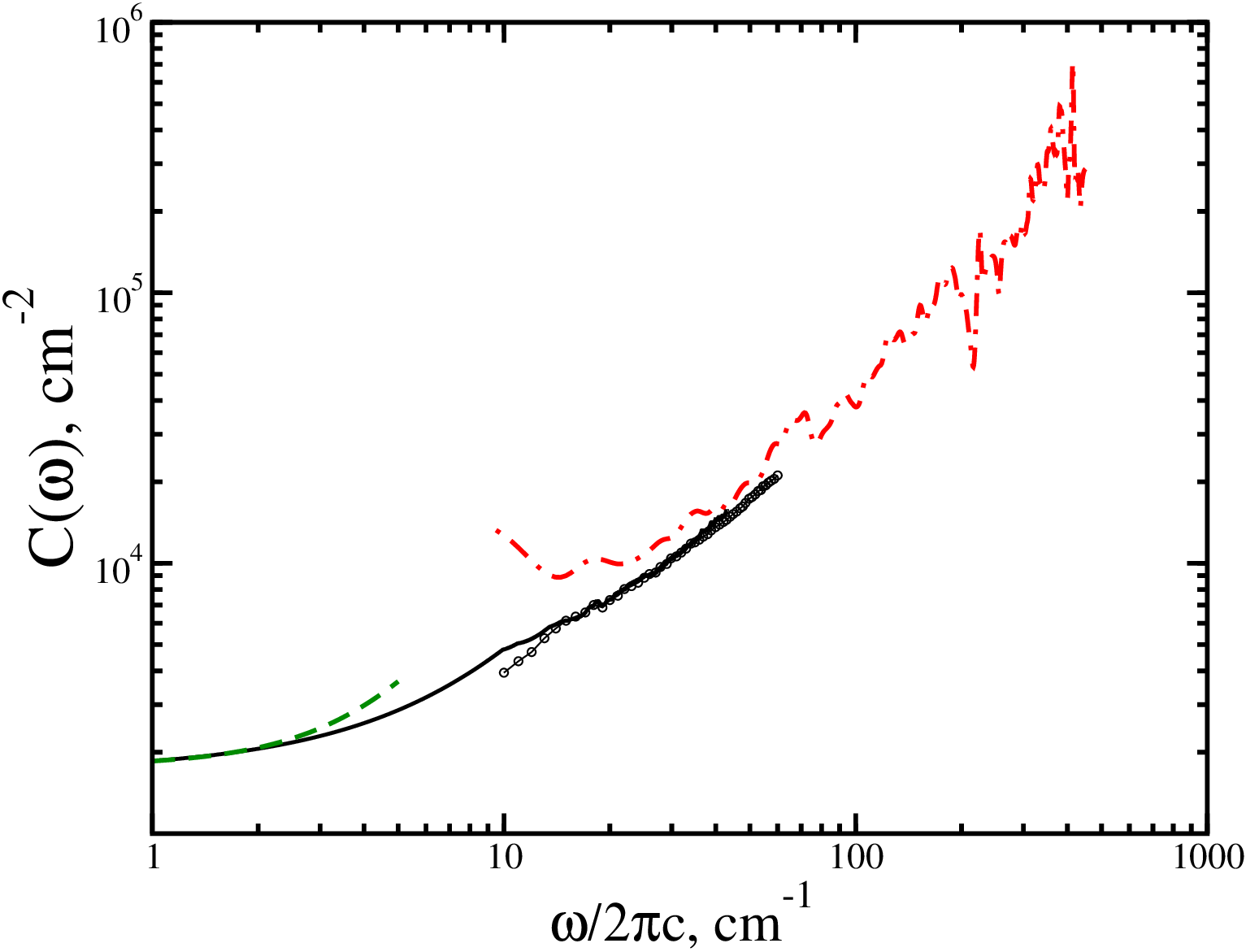}}   
\caption{(Color online)  
Experimental frequency dependence of the absorption  
coupling coefficient in the FIR range for As$_2$S$_3$ (solid line 
\cite{Taraskin_06:PRL} and circles \cite{Ohsaka_94}).  
The dot-dashed lines represent the numerical data obtained from a 
DFTB molecular-dynamics model. 
The dashed line shows the fit of the experimental data by Eq.~\ref{e0}  
with $A=1780~\text{cm}^{-2}$ and $B=75~\text{cm}^{-1}$.  
  }   
\label{f1}  
\end{figure}

As follows from the previous section, the coupling coefficient for the 
FIR  
absorption has a universal frequency dependence (see Eq.(\ref{e0})),  
containing a frequency-independent part and a frequency-dependent 
contribution ($\propto \omega^2$).  
The frequency dependence of the absorption coefficient, 
$\alpha(\omega)$, can be measured experimentally using THz 
time-domain spectroscopy \cite{Grischkowsky_90,Taraskin_06:PRL}.  
The VDOS can also be measured experimentally using, e.g. inelastic 
neutron scattering \cite{Fabiani_05,Isakov_93}.  
Therefore, the frequency dependence of the coupling coefficient for 
the FIR absorption, $C(\omega)=\alpha(\omega)/g(\omega)$, can be  
found  experimentally (see Fig.~\ref{f1}).  
Fitting experimental data with theory~(\ref{e0}) allows 
the values of constants $A$ and $B$ entering Eq.(\ref{e0}) to be 
estimated. 
We have done such a fitting to the experimental data 
for As$_2$S$_3$ glass (cf. the solid and dashed lines in Fig.~\ref{f1})  
and found the values of these constants to be   
$A=1780~\text{cm}^{-2}$ and  $B=75~\text{cm}^{-1}$ 
\cite{Taraskin_06:PRL}.  
 
The constant $B$ depends on the  structural characteristics of glass in  quite 
 a complicated fashion (see Eq.~(\ref{eq:S2b}) and cannot be used  
straightforwardly for extracting the relevant charge distributions.  
However, the constant $A$ is directly related to the width of the 
uncorrelated charge distribution,  
\begin{equation} 
\overline{\sigma}^2_1=\frac{A\overline{m}}{C_0} 
~,        
\label{eq:sigma1}  
\end{equation} 
where the coefficient $C_0$ is given by Eq.~(\ref{eq:C0}).  
Using Eq.~(\ref{eq:sigma1}),  
we have 
estimated the value of $\overline{\sigma}_1 \simeq 0.12~e$ for a-As$_2$S$_3$.  
 
\begin{figure}[ht] %
\vskip2truecm  
\centerline{\includegraphics[width=13cm]{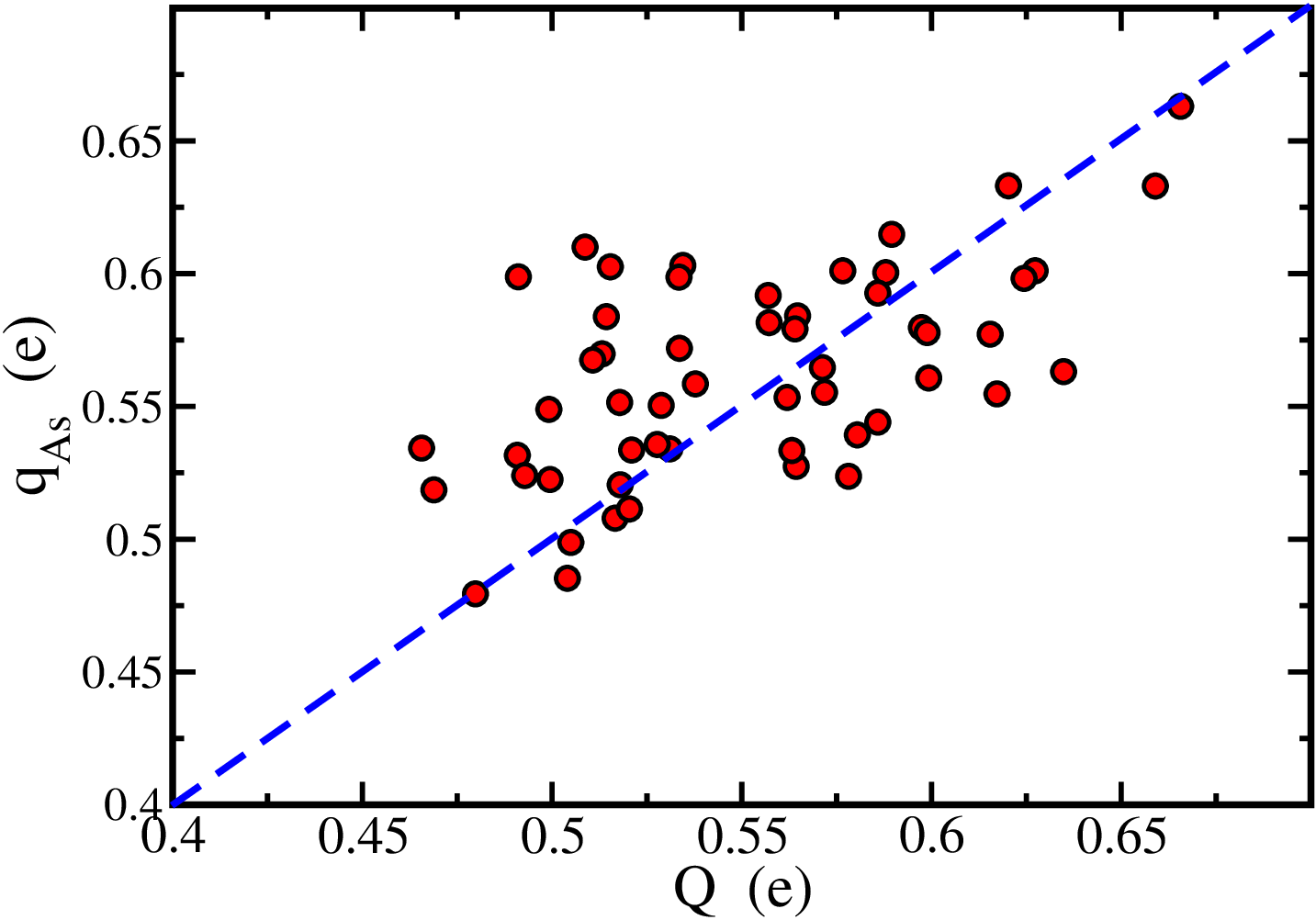}}   
\caption{(Color online)  
The charge of As atoms, $q_{\text{As}}$, in electron charge units 
versus the neutralising charge $Q=|\sum_{i\in\text{n.n.}}
q_{\text{S}}|/2$,  
where the summation is taken over all three nearest-neighbour sulfur  
atoms.  
The dashed line corresponds to the exact charge neutrality, 
$q_{\text{As}}=Q$, within the structural units.  
  }   
\label{f2}  
\end{figure}

In order to verify and support  the consistency of the presented model 
for the frequency dependence of the coupling constant, we have created 
a model of  As$_2$S$_3$ glass using  density functional theory-based  
tight-binding (DFTB) \cite{Porezag_95,Elstner_98}  
molecular-dynamics simulation \cite{Simdyankin_04}.  
First, we checked the local charge neutrality within AsS$_3$ 
pyramids, these being typical structural units in As$_2$S$_3$ glass.  
The results presented in Fig.~\ref{f2} clearly demonstrate 
correlations between the charge at the central As atom and 
surrounding S atoms.  
As expected for such a covalent material, the local charge neutrality 
is not exact and the data points fluctuate around the bisector due to 
contributions from $q_{1i}$.  
 
\begin{figure}[ht] %
\vskip2truecm  
\centerline{\includegraphics[width=13cm]{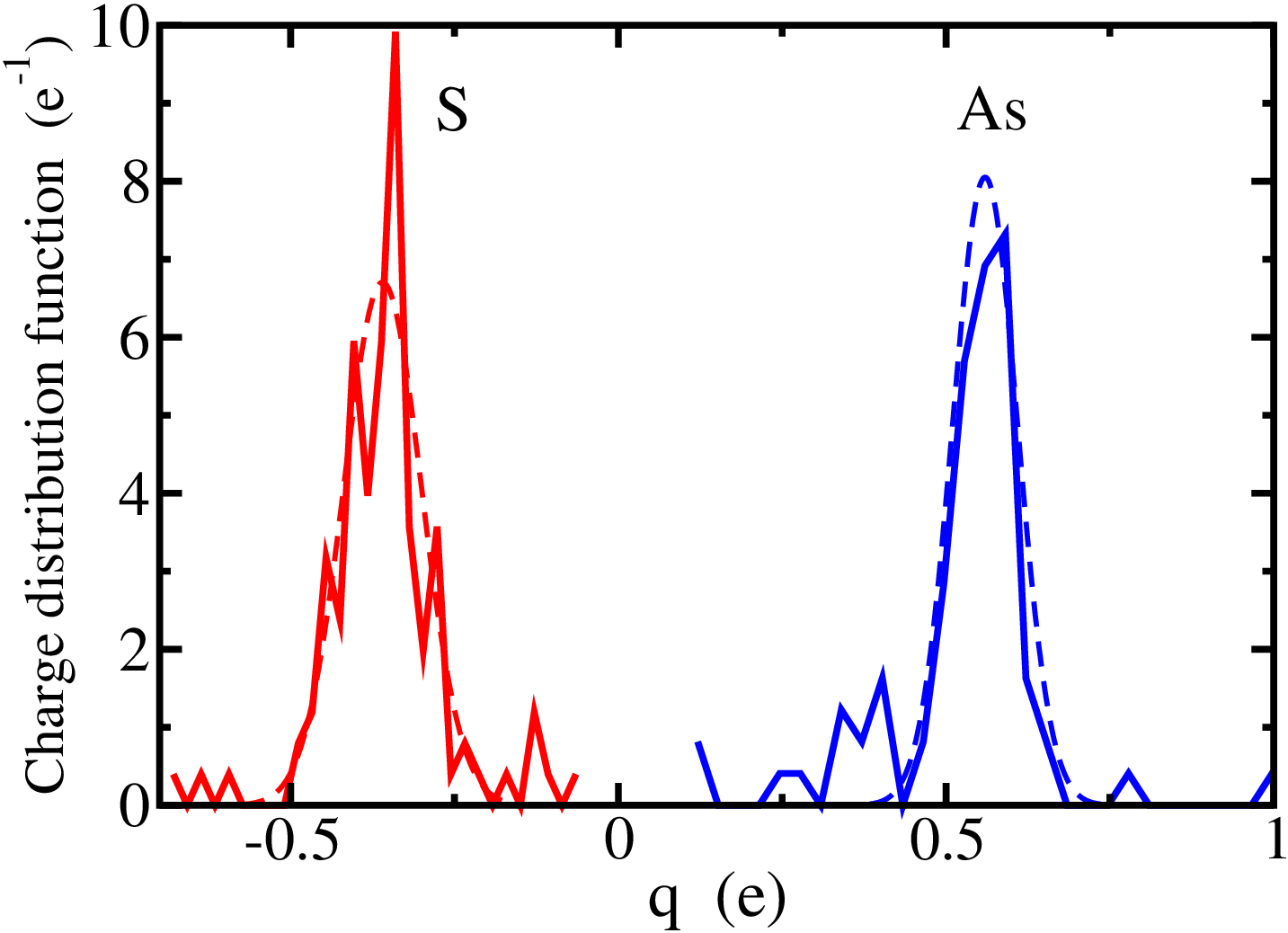}}   
\caption{(Color online)  
The atomic charge distributions for the DFTB model of As$_2$S$_3$.  
The left curve is for S atoms while the right one is for As atoms.  
The dashed curves show the Gaussian fits for these distributions 
with the mean values  $\overline{q}_{\text{As}} \simeq 0.56 $,  
$\overline{q}_{\text{S}} \simeq -0.36 $   and standard deviations  
$\sigma_{1\text{As}}\simeq 0.06$ and $\sigma_{1\text{S}}\simeq 0.05$.   
  }   
\label{f3}  
\end{figure}   
 
Second, we calculated the Mulliken charge distributions for As 
and S atoms in our DFTB model of As$_2$S$_3$ glass (see Fig.~\ref{f3}).  
The charges are distributed approximately normally (see the dashed lines in 
Fig.~\ref{f3})) around the mean values $\overline{q}_{\text{As}} \simeq 0.56 $ 
and $\overline{q}_{\text{S}} \simeq -0.36 $,   with the standard deviations  
$\sigma_{1\text{As}}\simeq 0.06$ and $\sigma_{1\text{S}}\simeq 0.05$.  
Therefore, the value of $\overline{\sigma}$ can be estimated as  
$\overline{\sigma}_1 = \sqrt{ (2/5)\sigma^2_{1\text{As}} +  
(3/5) \sigma^2_{1\text{As}}} \simeq 0.054$. 
This value is somewhat less than that estimated from the fit of the 
experimental data for $C(\omega)$ by Eq.~(\ref{e0}) (cf. the solid and dashed 
lines in Fig.~\ref{f1}), i.e.  
$\overline{\sigma}_1 \simeq 0.12$.  
Several possible effects could account for such a discrepancy. 
The Mulliken charges are an artificial way of assigning charge values 
to particular atoms within
the DFTB scheme - a different population analysis may give different
absolute values of the 
charges and thus different widths of the charge distributions. 
Another possible reason for the 
discrepancy is due to the experimental uncertainty in measuring the
absorption coefficient 
at very low frequencies,  
$\omega \alt 5 \text{cm}^{-1}$, caused by parasitic secondary 
reflections of THz pulses of light which, after the Fourier transform,   
can contribute in this frequency range.

\section{Conclusions}  
\label{s3} 
 
To conclude, it has been demonstrated that the frequency dependence of the 
coupling coefficient for far-infrared absorption can be used to
extract  
characteristics of the atomic  charge distribution in glasses.  
Namely, a fit of such an experimentally measured dependence by the theoretically 
predicted law, $A+B\omega^2$, allows the constants $A$ and $B$ to be 
extracted. The value of $A$ is simply proportional to the variance of the 
uncorrelated charge distribution, $\overline{\sigma}^2_1$.   
In the case of As$_2$S$_3$ glass, we  have estimated this value to be  
$\overline{\sigma}_1 \simeq 0.12$.  
A similar estimate of $\overline{\sigma}_1 \simeq 0.054$ has been obtained  
from a first-principles molecular-dynamics model of the same glass.

\end{document}